\documentclass[aps,pra,preprint,showpacs,floatfix,groupedaddress,superscriptaddress,longbibliography]{revtex4-1}

\usepackage[english]{babel}
\usepackage{float}
\usepackage{graphicx}
\usepackage[caption=false]{subfig}
\usepackage{amsmath}
\usepackage{amsfonts}
\usepackage{amssymb}
\usepackage{mathtools}
\usepackage{amsthm,enumitem}

\begin{document}

\title{Many-Body Effects in Models with Superexponential Interactions}

 \author{Peter Schmelcher}
  \email{Peter.Schmelcher@physnet.uni-hamburg.de}
 \affiliation{Zentrum f\"ur Optische Quantentechnologien, Universit\"at Hamburg, Luruper Chaussee 149, 22761 Hamburg, Germany}
 \affiliation{The Hamburg Centre for Ultrafast Imaging, Universit\"at Hamburg, Luruper Chaussee 149, 22761 Hamburg, Germany}

\date{\today}

\begin{abstract}
Superexponential systems are characterized by a potential where dynamical degrees of freedom
appear in both the base and the exponent of a power law.
We explore the scattering dynamics of many-body systems governed by superexponential potentials.
Each potential term exhibits a characteristic crossover via two saddle points from a region with a confining channel 
to two regions of asymptotically free motion. With increasing scattering energy in the channel we observe
a transition from a direct backscattering behaviour to multiple backscattering
and recollision events in this channel. We analyze this transition in detail by exploring both the properties
of individual many-body trajectories and of large statistical ensembles of trajectories.
The recollision trajectories occur for energies below and above the saddle points and typically
exhibit an intermittent oscillatory behaviour with strongly varying amplitudes.
In case of statistical ensembles the distribution of reflection times into the channel
changes with increasing energy from a two-plateau structure to a single broad
asymmetric peak structure. This can be understood by analyzing the corresponding momentum-time maps
which undergo a transition from a two-valued curve to a broad distribution. We close by providing
an outlook onto future perspectives of these uncommon model systems.
\end{abstract}

\maketitle

\section{Motivation and Introduction} 
\label{sec:introduction}

The interaction between the building blocks of matter typically
involve potentials with a power law dependence. For charged particles
this is the long-range Coulomb potential ($\propto \frac{1}{r}$) \cite{Jackson} whereas for neutral
constituents such as atoms \cite{Friedrich} or molecules \cite{Stone} their interaction 
at large distances can be of permanent dipolar character 
($\propto \frac{1}{r^3}$) or of induced dipolar origin, i.e. van der Waals interaction
($\propto \frac{1}{r^6}$). The importance of these interaction potentials is closely connected
to the fact that they describe the forces occuring in nature. This 
allows us to understand the structures and properties as well as
dynamics of few- to many-body systems via a bottom-up approach.

Complementary to the above
the development and analysis of more abstract models of interacting few- and many-body
systems possesses a rich history. These models are motivated, for
example, by the request for a thorough understanding of integrability versus
nonintegrability \cite{Sirker,Giamarchi}, the mechanisms of the transition from few- to many-body systems 
\cite{Zinner}, and the emergence of thermodynamical behaviour in the particle number to
infinity limit \cite{Samaj}. A particularly striking and impactful paradigm 
is a system of contact interacting particles in one spatial dimension for
which the interaction among the particles is contracted to a single point
providing corresponding boundary conditions. This leads to an intricate
relationship between impenetrable bosons and fermions in one dimension
\cite{Busch,Girardeau}. After many years of their discovery and investigation,
these models are nowadays used extensively to describe the physics of ultracold
quantum gases and Bose-Einstein condensates \cite{Pethick}. Due to the separation
of length scales in dilute gases for which the range of the collisional interactions
is typically much smaller than the distance between the particles as well as the overall
size of the atomic cloud the model of contact interacting atoms provides a valid
description of the structure and properties as well as dynamics of these many-body 
systems \cite{Pethick,Pitaevskii}.

While many naturally occuring interactions involve power law potentials with
a constant exponent, the properties and dynamics of models with so-called superexponential
interactions have been explored very recently \cite{Schmelcher1,Schmelcher2,Schmelcher3,Schmelcher4}.
Rendering the exponent time-dependent one arrives at a periodically driven power-law oscillator \cite{Schmelcher1}.
Covering weak and strong confinement during a single driving period, the resulting classical phase
space comprises not only regular and chaotic bounded motion but exhibits also a tunable exponential
Fermi acceleration. Note that the fundamental mechanisms of exponential acceleration
and their applications have come into the focus of research in nonlinear dynamics
in the past ten years \cite{Turaev1,Shah1,Liebchen,Shah2,Turaev2,Turaev3,Batistic,Pereira}. 
A major step forward in the direction of superexponential dynamics is provided by the
so-called superexponential self-interacting oscillator \cite{Schmelcher2}. The potential
of this oscillator takes on the unusual form $V = |q|^q$ where the exponent
depends on the spatial coordinate $q$ of the oscillator. The exponentially varying nonlinearity
leads to a crossover in the period of the oscillator from a linearly decreasing to a nonlinearly
increasing behaviour with increasing energy. This oscillator potential possesses
a hierarchy of (derivative) singularities at its transition point $q=0$ which are responsible for this crossover
and lead to a focusing of trajectories in phase space. The spectral and eigenstate properties
of the corresponding quantum superexponential oscillator \cite{Schmelcher3} do reflect this
transition equally: the ground state shows a metamorphosis of decentering, asymmetrical squeezing
and emergence of a tail. Signatures of the crossover can be seen in the excited states by analyzing
e.g. their central moments which show a transition from an exponentially decaying to an increasing
behaviour. 

A major step forward on the route to superexponentially interacting many-body systems is represented
by the very recently explored two-body case \cite{Schmelcher4}. The latter represents a fundamental
building block for many-body systems and is therefore a key ingredient to the present work. The
underlying Hamiltonian contains the superexponential interaction potential $V = |q_2|^{q_1}$
which couples the degrees of freedom $q_1$ and $q_2$ in an exponential manner. The resulting potential
landscape exhibits two distinct regions: a region where motion takes place in a confining channel (CC) with
varying transversal anharmonicity and a region with asymptotically free motion. These regions are connected
via two saddle points allowing for a deconfinement transition between the confined and free motion.
In ref.\cite{Schmelcher4} the dynamics and in particular scattering functions have been analyzed in
depth for this peculiar interaction potential thereby demonstrating the impact of the dynamically 
varying nonlinearity on the scattering properties.

On basis of our understanding gained for the fundamental two-body system it is now
a natural next step to investigate many-body superexponentially interacting systems
which we shall pursue here. We thereby focus on systems with a single exponent ($q_1$)
and many base ($q_i, i=2,...,N$) degrees of freedom with interaction terms of the form $\propto |q_i|^{q_1}$.
We provide a comprehensive study of the many-body scattering dynamics thereby analyzing 
the mechanisms of the collisional dynamics in the CC with increasing energy.
Due to the presence of many transversal channel degrees of freedom 
a plethora of energy transfer processes are enabled. As a consequence the incoming
longitudinal $q_1$ scattering motion undergoes in the low-energy regime
a transition from a step-like to a smooth behaviour.
While the two-body scattering allows for energies below the saddle points only for a monotonous
behaviour of the incoming and outgoing motion we show that many-body processes
lead to an intricate combination of backscattering and recollision events. This includes a
highly oscillatory behaviour with multiple turning points emanating from
the saddle point region and reaching out into the CC. This oscillatory
and intermittent scattering motion exhibits largely fluctuating amplitudes, a feature which is
absent in the case of two-body scattering.
Our analysis comprises the energy-dependent behaviour of individual trajectories as well as the
statistical behaviour of ensembles including an analysis via momentum-time and turning point maps.

This work is structured as follows. In section \ref{sec:hamiltonian} we introduce the 
Hamiltonian and discuss the underlying interaction potential landscape as well as the
classification of the dynamics in terms of invariant subspaces. Section \ref{sec:dynamics1}
contains a detailed discussion of the individual many-body trajectories in the low-, intermediate and
high energy regime. Section \ref{sec:dynamics2} provides an analysis of the statistical
ensemble properties with a focus on the reflection time distribution. Section \ref{sec:sac} presents
our summary and conclusions including a brief outlook.

\section{The Superexponential Hamiltonian and Potential Landscape} 
\label{sec:hamiltonian}

This section is dedicated to the introduction of the Hamiltonian and a discussion
of the landscape of its interaction potential. We will also provide 
the invariant subspaces of the dynamics. Our superexponential Hamiltonian 
takes on the following appearance

\begin{equation}
{\cal{H}} = {\cal{T}} + {\cal{V}} = \sum_{i=1}^{N} \frac{p_i^2}{2} + \sum_{k=2}^{N} |q_k|^{q_1}
\label{eq:hamiltonian1}
\end{equation}

where ($q_i,p_i, i=1,...,N$) are the canonically conjugate
coordinates and momenta of our 'effective particles or entities' respectively,
and in this sense we refer to the above Hamiltonian as a many-body Hamiltonian. Equally the term
'interaction' is employed here to indicate that the individual potential terms $\propto |q_k|^{q_1}$ 
depend on two particles coordinates. Note that both the base degrees of
freedom (dof) ($q_k,k=2,...,N$) as well as the exponent dof $q_1$ possess a corresponding
kinetic energy and therefore evolve dynamically. The individual $N-1$ interaction terms $|q_k|^{q_1}$ share
a single exponent dof and therefore the interaction between the dof $q_k$ takes place indirectly
via the dof $q_1$. In other words the dof $q_1$ could be seen as a common dof shared by all
the base dof ($q_k,k=2,...,N$). The superexponential potential ${\cal{V}}= \sum_{k=2}^{N} |q_k|^{q_1}$
(SEP) mediates the interaction among the dof $q_k, k=1,...,N$. 
Obviously, the above model Hamiltonian does not exhibit well-established symmetries
such as a translation invariance. It possesses however an exchange symmetry with respect to
the base dof $q_k,k=2,...,N$ since they all couple in the same manner to the exponent dof $q_1$. This will
allow us to conclude upon invariant dynamical subspaces (see below). We remark that
the Hamiltonian \ref{eq:hamiltonian1} is a specific choice out of many possible superexponential Hamiltonians
(see discussion in the conclusions section \ref{sec:sac}) which is motivated by the appearance
of only a single exponent dof which promises a more straightforward interpretation of the resulting
many-body dynamics.

\begin{figure}
\parbox{15cm}{\includegraphics[width=14cm,height=8cm]{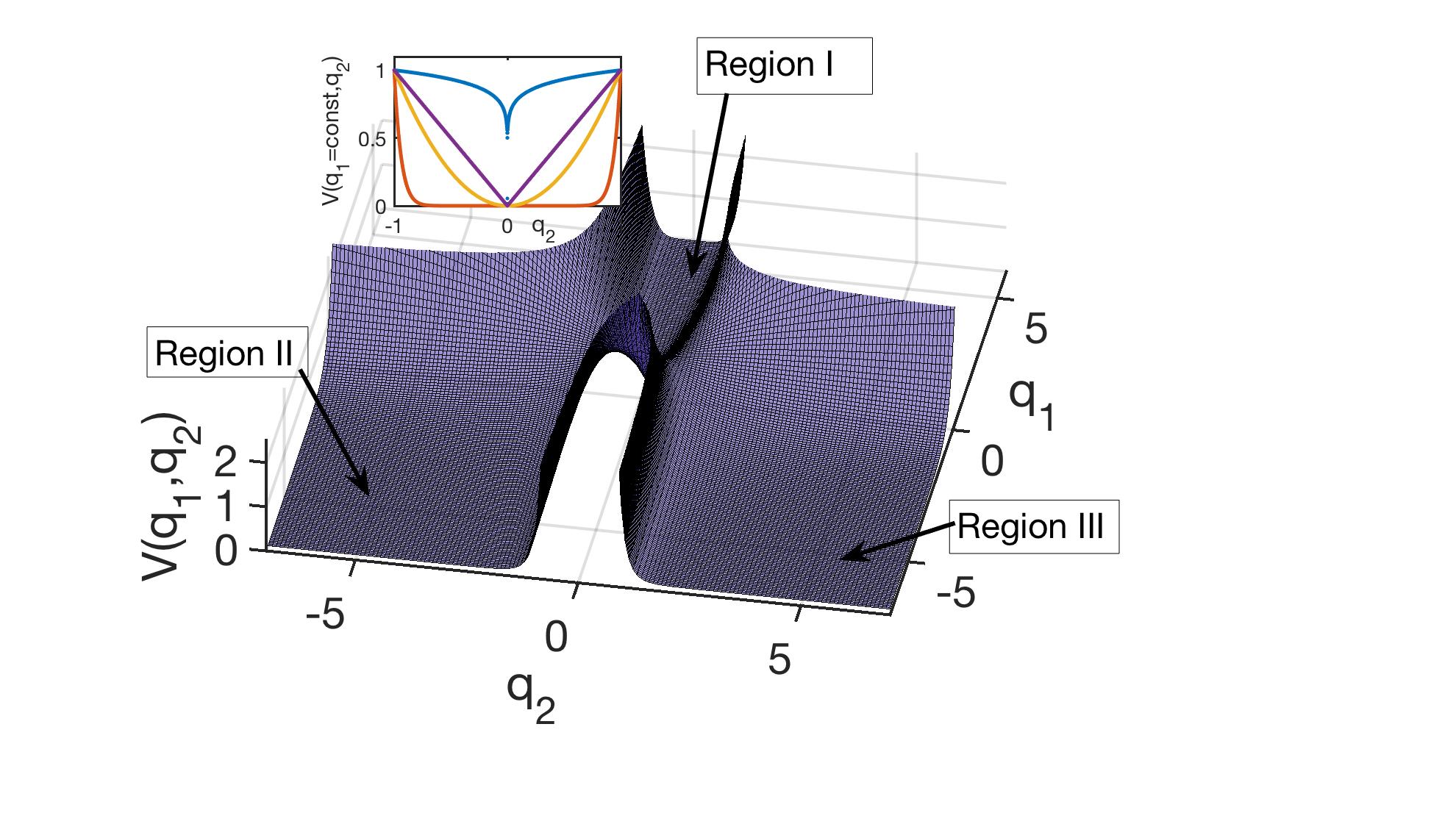}} 
\caption{The potential energy landscape of a single interaction term 
$V(q_1,q_2)=|q_2|^{q_1}$. The CC
(region I) as well as the two regions of asymptotically free motion, regions II and III, are indicated.
The inset shows intersections of the potential energy along the $q_2$ coordinate i.e. 
$V(q_1=\mathrm{const},q_2)$ for $q_1=0.1,1,2,16$ corresponding to the curves from top to bottom.}
\label{fig1}
\end{figure}

In ref.\cite{Schmelcher4} the superexponentially interacting
two-body system containing a single interaction term has been explored and analyzed in detail. 
To be independent and to set the stage for the many-body case we will in the following briefly summarize the
main properties of the potential landscape for a single interaction term $|q_2|^{q_1}$. It shows
(see Figure \ref{fig1}) for $q_1 > 0$ (region I) a CC leading to a bounded motion w.r.t. the
coordinate $q_2$ and an unbounded motion for the dof $q_1$. The transversal confinement of this channel 
illustrated by the intersection curves $V(q_1=\mathrm{const},q_2)$ (see inset of Figure \ref{fig1})
continuously changes with increasing values of $q_1$: the cusp for $q_1 < 1$ turns into a linear confinement
for $q_1 = 1$, a quadratic one for $q_1=2$ and finally into a steep wall anharmonic confinement for $q_1 \gg 2$.
For $q_1 \rightarrow \infty$ the channel confinement is that of a box with infinite walls.

The channel region I is connected via two saddle points at energy $E=1$ to regions II and III which
exhibit asymptotically ($q_1 \rightarrow -\infty, q_2 \rightarrow \pm \infty$) free motion.
Regions II and III are separated by a repulsive potential barrier with a (singular) maximum at $q_2=0$. 
In region II both particles move in a correlated manner in the same direction ($p_1,p_2<0$)
and in region III they move in opposite directions ($p_1<0,p_2>0$). To conclude, while the appearance
of the SEP is very simple it shows an interesting geometrical structure. Let us now turn back
to the many-body problem.

The SEP ${\cal{V}}(q_1,q_2,...,q_N)=\sum_{k=2}^{N} |q_k|^{q_1}$
in eq.(\ref{eq:hamiltonian1}) possesses stationary points, i.e.
zero derivatives $\frac{\partial {\cal{V}}}{\partial q_i}=0, \forall i=1,...,N$, 
at the positions $q_1=0,q_i=\pm 1, i=2,...,N$. The resulting Hessian possesses a zero determinant, but a 
more detailed analysis shows that the extrema have unstable and stable directions, i.e. they are saddle
points. The energies of the extrema are $E={\cal{V}}(q_1=0,\{q_i=\pm 1 \} )=(N-1)$.

Since the Hamiltonian equations of motion belonging to the Hamiltonian (\ref{eq:hamiltonian1})
possess a singularity for $q_1<0,q_i=0,i=2,...,N$, we introduce a regularization parameter
$\epsilon > 0$ for the SEP which now reads ${\cal{V}}_{reg}(q_1,q_2,...,q_N;\epsilon)=\sum_{k=2}^{N}
(\sqrt{q_k^2+\epsilon})^{q_1}$. This facilitates the numerical integration of the corresponding
Hamiltonian equations of motion which read

\begin{eqnarray}
{\dot{q_i}} = p_i \hspace*{1cm} i=1,...,N \label{eq:heom1}\\
{\dot{p}}_1 = - \sum_{k=2}^{N} \left(\sqrt{q_k^2+\epsilon} \right)^{q_1} \ln \left( \sqrt{q_k^2+\epsilon} \right)
\label{eq:heom2}\\
{\dot{p}}_i = - \left(\sqrt{q_i^2+\epsilon} \right)^{q_1-2} q_1 q_i \hspace*{1cm} i=2,...,N \label{eq:heom3}
\end{eqnarray}

Typical values chosen for the numerical simulations are $\epsilon = 10^{-8}$.
The equation of motion for $p_1(t)$ depends symmetrically on all $q_k, k=2,...,N$ due to the above-mentioned
exchange symmetry. Note the appearance of the logarithm which will be of major importance
for the later on observed dynamics. The equation of motion of $p_i(t),i=2,...,N$ 
depends only on the coordinates $q_i$ and $q_1$ and these equations ($i=2,...,N$) are structurally invariant
due to the exchange symmetry. This means, that for equal initial conditions (ICs) of all $(q_i,p_i),i=2,...,N$
at $t=t_0=0$ the dynamics of all $q_i(t),i=2,...,N$ will be identical.

Let us elaborate on this in some more detail since it allows us to identify a hierarchy of invariant
subspaces that classify the dynamics. The exchange symmetry among the $N-1$ particles with respective coordinates
and momenta $(q_i,p_i),i=2,...,N$ can be either (i) completely broken (ii) partially broken or (iii) fully
maintained by the corresponding ICs. We therefore partition the complete phase space
of ICs into subspaces as follows. We divide the $2N-2$-dimensional total phase space ${\cal{P}}$
of the dof $q_i,i=2,...,N$ 
into dynamically invariant subspaces ${\cal{C}}_i$ of identical ICs which
lead consequently to an identical dynamics (trajectories).
Here the invariance refers to the exchange of (initial) phase space coordinates in the corresponding
subspace ${\cal{C}}_i$. These subspaces represent a classification of the dynamics.
More specifically, we define a series of positive integers $\{n_i\}= n_1,....,n_k$ with $\sum_{i=1}^{k}n_i=(N-1)$
where $n_i$ is the maximal dimension of the subspace ${\cal{C}}_i$ with identical initial phase space coordinates.
A complete set of ICs (and resulting trajectories) is then given by the 
decomposition $\cup_{i=1}^{k} {\cal{C}}_i = {\cal{P}}$. This set involves, per definition, $k$ different
classes of identical trajectories, the $i-th$ class containing $n_i$ identical phase space coordinates.
A remark concerning the resulting combinatorics is in order.
For a single subset of $l$ identical ICs only there is $\binom{N-1}{l}$ 
possible configurations or subspaces with $l \le (N-1)$. For $r$ subsets each one with $k$ identical ICs
the number of possibilities is $\sum_{i=0}^{r-1} \binom{N-ik-1}{k}$. This generalizes to
the case of an arbitrary number of subspaces of properly chosen dimensions with identical ICs.

\section{Dynamics: Individual Many-Body Trajectories}
\label{sec:dynamics1}

This section is devoted to the exploration of the many-body dynamics by analyzing individual
trajectories which illustrate the relevant collisional processes. We note that these trajectories
are representative and show the typical observed behaviour.
The general procedure is as follows. We will simulate the dynamics in the CC (region I) for
incoming ($p_1 <0$) trajectories starting at $t=0$ in the outer part of the channel at $q_1=30$.
At this value of $q_1$ the transverse profile of the channel represented by the intersections of the individual
interaction potential terms ${\cal{V}}(q_1=\mathrm{const},q_2)$ is already very similar to a box confinement.
We will then study the dynamics with increasing total energy and for different subspaces of identical
ICs.

\subsection{Low energy scattering} \label{dyn:lse}

Let us start by assuming that all ICs of the coordinates and momenta $(q_i,p_i),i=2,...,N$ are identical.
Since then (see discussion in section \ref{sec:hamiltonian}) all dynamical evolutions $q_i(t),p_i(t)$ are
identical this case is similar to the case of the corresponding superexponentially interacting two-body 
system \cite{Schmelcher4}. Let us summarize the main features and characteristics of the 
dynamics for the two-body case (where the total potential reads ${\cal{V}}=|q_2|^{q_1}$)
for reasons of comparison to the actual many-body case. Since the exponent dof $q_1$
provides the confinement for the dof $q_2$ the time evolution of $q_2(t)$ shows bounded oscillations
in the channel (see Figure \ref{fig3}(a) for a specific case of the many-body system).
For large values of $q_1$ this confinement is strongly anharmonic and close to a box-like
confinement: as a consequence the channel is approximately flat for $-1 \lesssim q_2 \lesssim +1$ and energy exchange
processes (between particles but also from kinetic to potential energy for a single dof) happen only close to the
turning points of the $q_2$ oscillations. Opposite to this the $q_1$-motion is not oscillatory and is unbounded.
This can be argued as follows. Inspecting eq.(\ref{eq:heom2}) and specializing it to the case
of a single base dof $q_2$ one realizes that the r.h.s. is positive ($\epsilon = 0$) as long as the logarithm
is negative, which implies $q_2 < 1$. A necessary condition for ${\dot{p_1}} < 0$ to happen
is then given by the occurence of $q_2 > 1$ which implies that the total energy $E > 1$. The
latter is however the energy of the saddle points in the two-body problem. To conclude, this
means that for energies below the saddle point energies the two-body scattering in the
CC involves a time evolution $q_1(t)$ with exclusively ${\ddot{q}}_1>0$ i.e.
the incoming $q_1(t)$ trajectory possesses a single turning point ! As a consequence,
$q_1(t)$ cannot perform an oscillatory bounded motion but describes simply a direct in-out
scattering process finally escaping asymptotically to $q_1 \rightarrow \infty$. 
In this sense multiple scattering processes are not encountered and scattering is not
chaotic, i.e. there is even no transient dynamics with nonzero Lyapunov exponents.
This situation changes when considering the many-body situation. Here the saddle point
energy is given by $E_s=(N-1)$ and the dynamics of $(q_1,p_1)$ is determined (see 
eqs.(\ref{eq:heom1},\ref{eq:heom2})) by the sum over all forces involving the dof $q_i,i=2,...,N$.
This sum has to become overall positive (as a combination of the appearing logarithms and their
'above threshold' $q_k >1, k \in \{2,...,N\}$ arguments) in order to enable ${\dot{p_1}} < 0$
and to provide multiple turning points as well as an oscillatory dynamics: it is an inherent many-body process.

\begin{figure}
\hspace*{-6cm} \parbox{12cm}{\includegraphics[width=18cm,height=12cm]{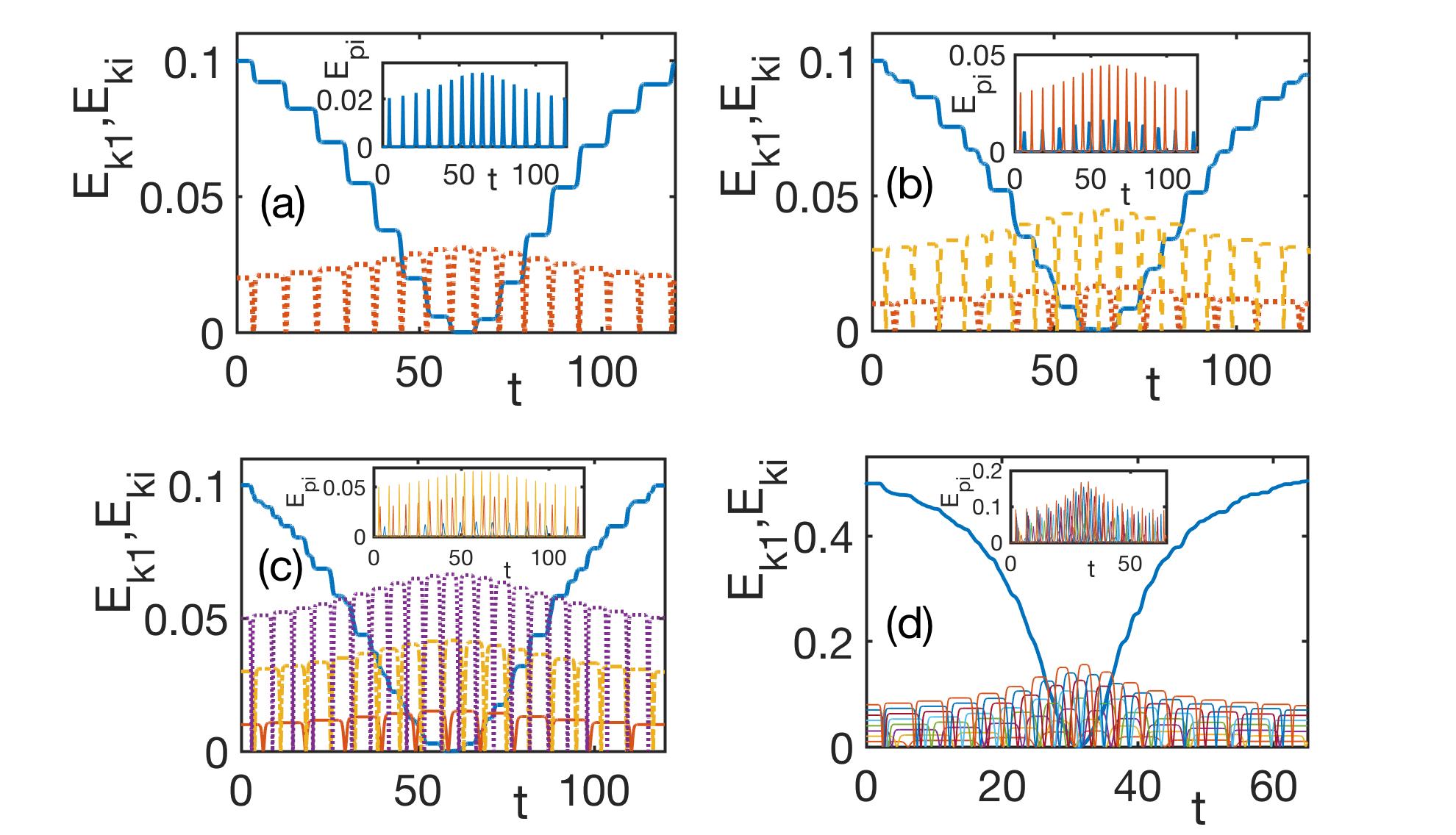}} 
\caption{The kinetic energies $E_{k1}$ (blue solid curve),
$E_{ki}$ (dotted, dashed and dot-dashed curves), and
potential energies $E_{pi}$ (see inset), belonging to the dof $q_1,q_i$, respectively, 
as a function of time $t$ for individual trajectories. Initial conditions are $q_1=30,q_i=0,i=2,...,N$.
(a) Total energy $E=0.28$ and initial conditions $E_{k1}=0.1,E_{ki}=0.02$. Note that all curves
$E_{ki},i=2,...,N$ are identical due to identical ICs. Similar statements hold for (b,c).
(b) Total energy $E=0.29$ and initial conditions $E_{k1}=0.1$, $E_{ki}=0.01, i=2-5$, $E_{kj}=0.03,j=6-10$. 
(c) Total energy $E=0.37$ and initial conditions $E_{k1}=0.1$, $E_{ki}=0.01, i=2-4$, $E_{kj}=0.03,j=5-7$,
$E_{kl}=0.05, l=8-10$, (d) Total energy $E=0.95$ and initial conditions $E_{k1}=0.5$, 
$E_{ki}=0.01 \cdot (i-1),i=2-10$. All simulations involve $N=10$ particles.}
\label{fig2}
\end{figure}

\begin{figure}
\parbox{8cm}{\includegraphics[width=7.6cm,height=6.6cm]{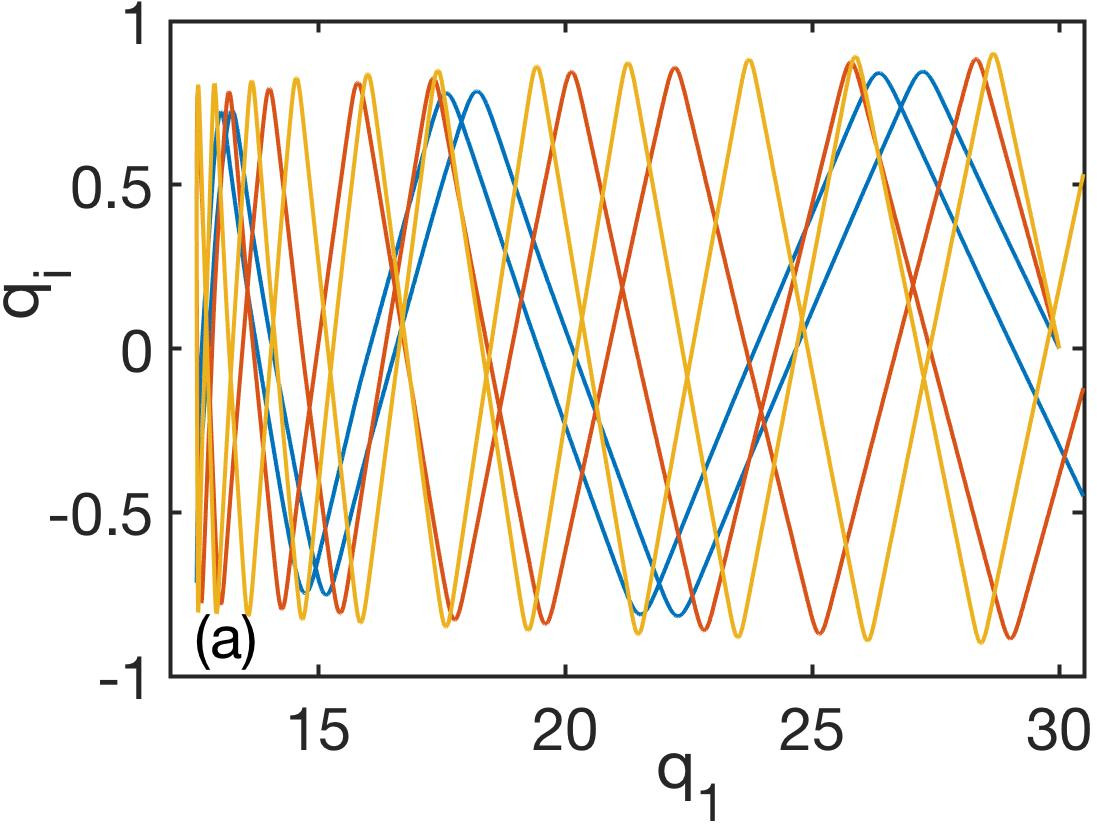}} 
\parbox{8cm}{\includegraphics[width=7.3cm,height=6.3cm]{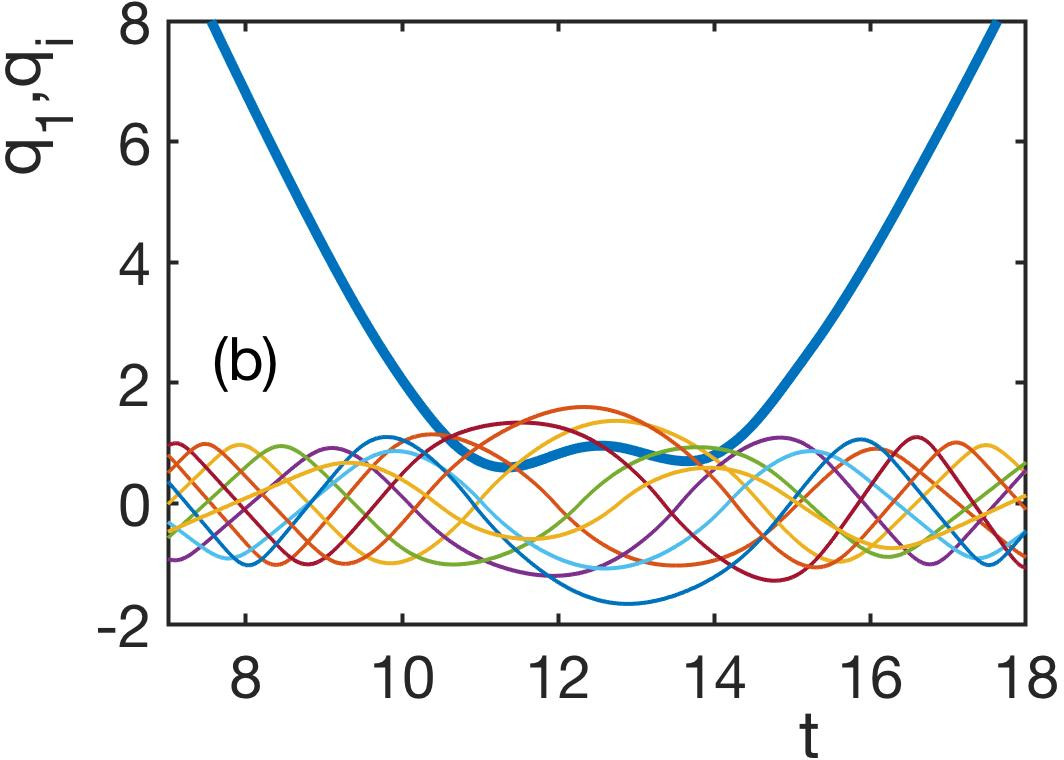}} 
\parbox{8cm}{\includegraphics[width=7.0cm,height=6.3cm]{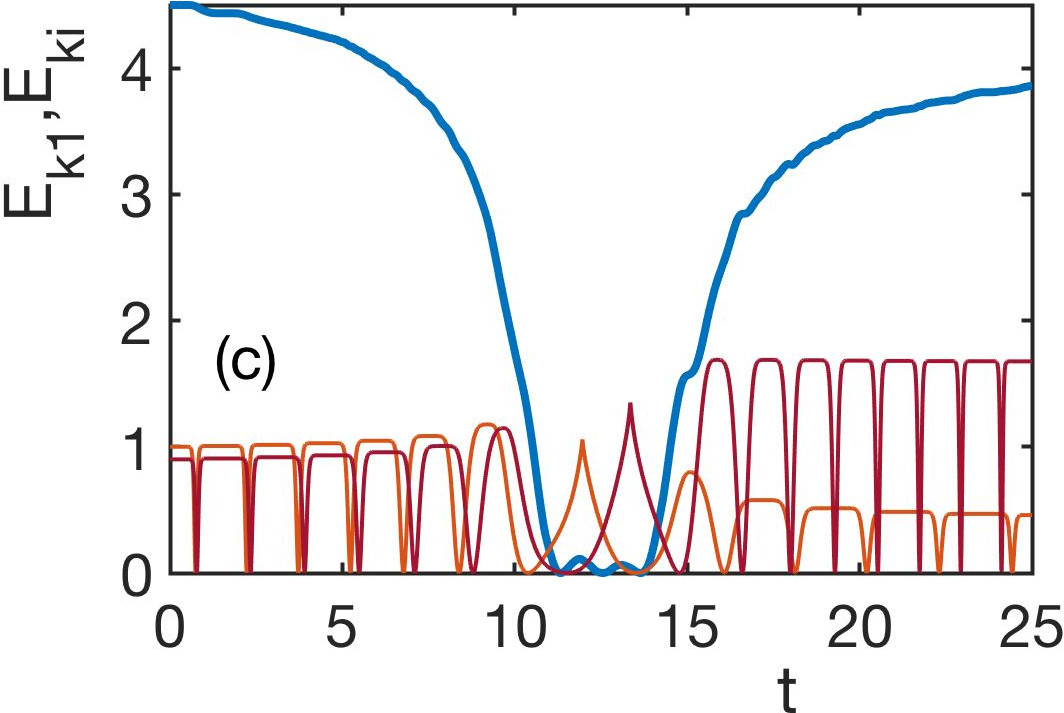}} 
\caption{(a) A $(q_1,q_i)$ graph of scattering trajectories in the CC with initial
conditions $q_1=30,q_i=0,i=2,...,N$ and $E_{k1}=0.1,(E_{ki}=0.01,i=2-4),(E_{kj}=0.03,j=5-7),(E_{kl}=0.05,l=8-10)$
and for a total energy $E=0.37$. Clearly visible are three types of transversal $q_i(t)$ oscillations and
the reflection process at the minimal value of $q_1$.
(b) Time evolution $q_1(t),(q_i(t),i=2,...,N)$ of a scattering trajectory with total energy $E=10.6$ via the 
CC. Initial conditions for the coordinates are the same as in (a), and $E_{ki}=4.5,1,0.4,0.7,0.9,1.1,
0.5,0.8,0.6,0.1$ corresponding to $i=1,...,10$. An oscillatory behaviour in the saddle point region is
clearly visible. (c) Same as in (b) concerning the parameters and ICs. Shown are the kinetic energies
$E_{k1}(t),E_{ki}(t)$ for a few selected particles to get a representative view. Multiple oscillations
and inelastic processes are evident.}
\label{fig3}
\end{figure}

Figure \ref{fig2}(a) shows the kinetic energies $E_{k1}=\frac{p_1^2}{2}$ and $E_{ki}=\frac{p_i^2}{2}$ as well as the
corresponding potential energies $E_{pi}=|q_i|^{q_1}$ (see inset) as a function of time for the
scattering process of a system of $N=10$ particles with the total energy $E=0.28$.
Here all ICs of the base dof are identical i.e. the particle exchange symmetry among the dof $q_i,i=2,...,N$
is fully maintained and their dynamics is the same. Therefore, we expect that the
above described properties of the two-body scattering dynamics should also appear here.

Indeed, the initial kinetic energy $E_{k1}(t=0)=0.1$ belonging to the subsequent time evolution $(q_1(t),p_1(t)$
decreases monotonically to zero and subsequently increases in the course of the scattering
process (see Figure \ref{fig2}(a)).
$E_{k1}(t)$ exhibits a sequence of plateaus which correspond (see discussion above) to the
traversal of $q_i(t)$ of the bottom of the CC, while the phases of rapid changes of $E_{k1}(t)$ between
two plateaus is caused by the dynamics in the vicinity of the potential walls. These facts 
are consistent with the behaviour of the potential energy $E_{pi}$ (see inset of Figure \ref{fig2}(a)) which exhibits
pronounced peaks during these collisions with the potential walls. For reasons of energy conservation
$E_{ki}(t)$ show then corresponding dips.

As a next step let us break the (total) exchange symmetry among the base dof by firstly inspecting the case of
two sets of identical ICs. Figure \ref{fig2}(b) shows the kinetic $E_{k1},E_{ki}$ and the potential energies
$E_{pi}$ for a total energy $E=0.29$ and ICs $(E_{ki}=0.01,i=2-5);(E_{kj}=0.03, j=6-10)$. Since we have now
two different sets of identical dynamics namely $(q_i(t),i=2-5);(q_j(t),j=6-10)$ a partial exchange symmetry
remains. The corresponding time evolution $E_{k1}(t)$ carries now the signatures of two different transversal motions
$(q_i(t),q_j(t))$: the overall decrease and subsequent increase due to the collision process exhibits now
a 'superposition' of plateau-like structures. Correspondingly, there is two different kinds of time evolution
of kinetic energies $E_{ki}(t),E_{kj}(t)$ which show sharp dips at the time instants where the kinetic 
energy $E_{k1}(t)$ varies rapidly in between two plateaus. The associated potential energies $E_{pi}(t),E_{pj}(t)$ 
(see inset of Figure \ref{fig2}(b)) show pronounced peaks at the time instants of collisions with
the potential walls which correspond to the time instants of the previously mentioned dips of $E_{ki}(t),E_{kj}(t)$.

Figure \ref{fig3}(a) shows the channel dynamics for the base and exponent dof $q_1,q_i,i=2,...,N$
and Figure \ref{fig2}(c) the corresponding kinetic and potential energies for a total energy $E=0.37$ 
for the case of three sets of identical ICs. The dynamics of $E_{k1}(t)$ shows a larger number of 
plateaus which, due to their partial overlap, gradually become washed-out.
This becomes even more pronounced for the case of no identical ICs and a total energy $E=0.95$ shown
in Figure \ref{fig2}(d): here the time evolution $E_{k1}(t)$ becomes almost smoothly decreasing
and subsequently increasing, i.e. without any pronounced plateau-like structures. In Figures \ref{fig2}(c,d)
the time evolutions of the kinetic energies $E_{ki}(t)$ show an increasing number of dips and
in case of the potential energies $E_{pi}(t)$ an increasing number of peak structures (see corresponding insets).
In Figure \ref{fig2}(d) there exists already a rather dense accumulation of peaks ($E_{pi}(t)$, see
inset) and dips ($E_{ki}(t)$) due to the
many collisions of the particles with dof $q_i(t)$ with the walls of the interaction potential ${\cal{V}}$.

\subsection{Intermediate energy scattering} \label{dyn:ise}

We remind the reader of the fact that the saddle point threshold energy
is $E_{s}=(N-1)$ which amounts to $E_s=9$ for our prototypical
10 particle system. As discussed above (see section \ref{dyn:lse}) the two-body case as well as the
case of identical ICs for all base dof (in the many particle case)
show only a single turning point, i.e. a direct in-out scattering
behaviour, for the exponent dof $q_1(t)$ for energies $E<E_s$.
This statement holds also for the low energy scattering $E \ll E_s$ discussed in the previous subsection
where a transition of the dynamics $E_{k1}(t)$ from plateau-dominated to a smooth behaviour has been
observed with increasing number of different ICs.

\begin{figure}
\parbox{8cm}{\includegraphics[width=7.6cm,height=6.6cm]{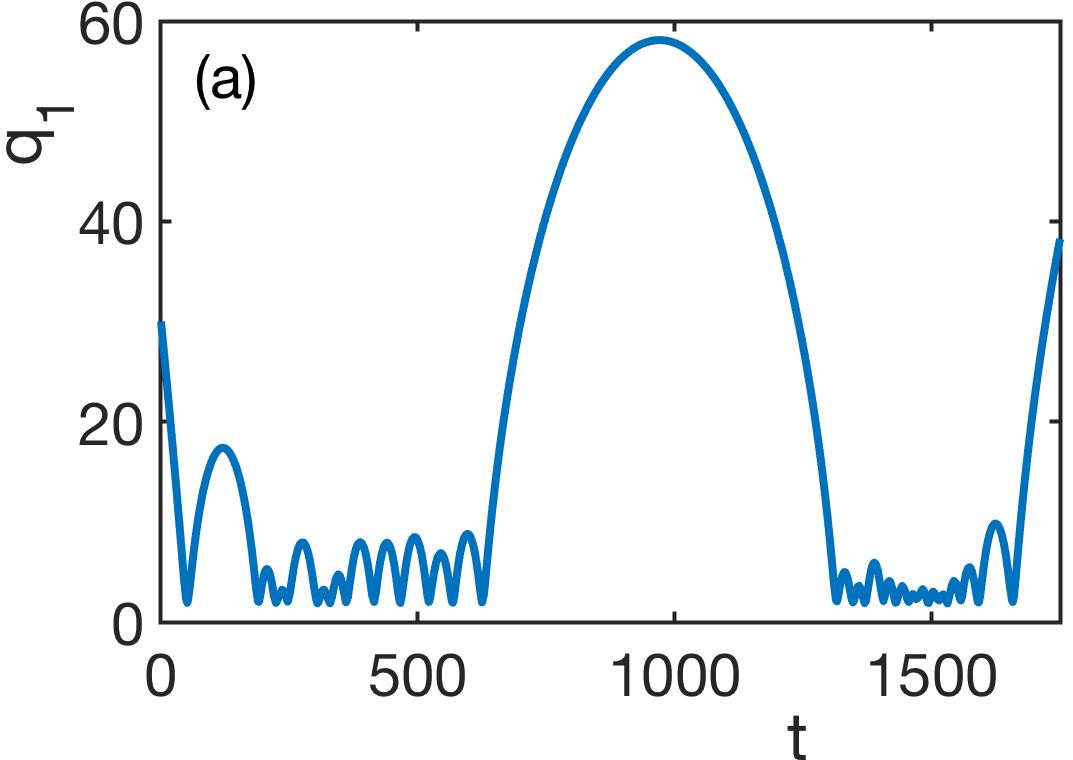}} 
\parbox{8cm}{\includegraphics[width=8.2cm,height=6.5cm]{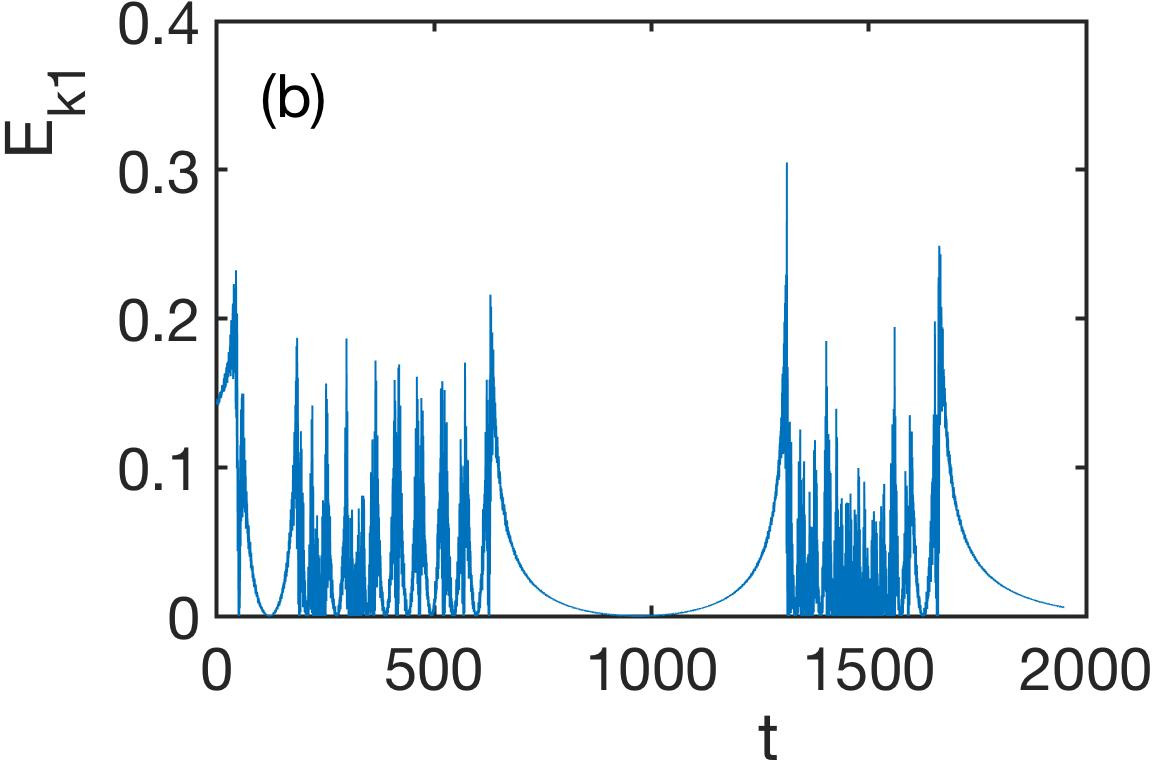}} 
\parbox{8cm}{\includegraphics[width=8.0cm,height=6.3cm]{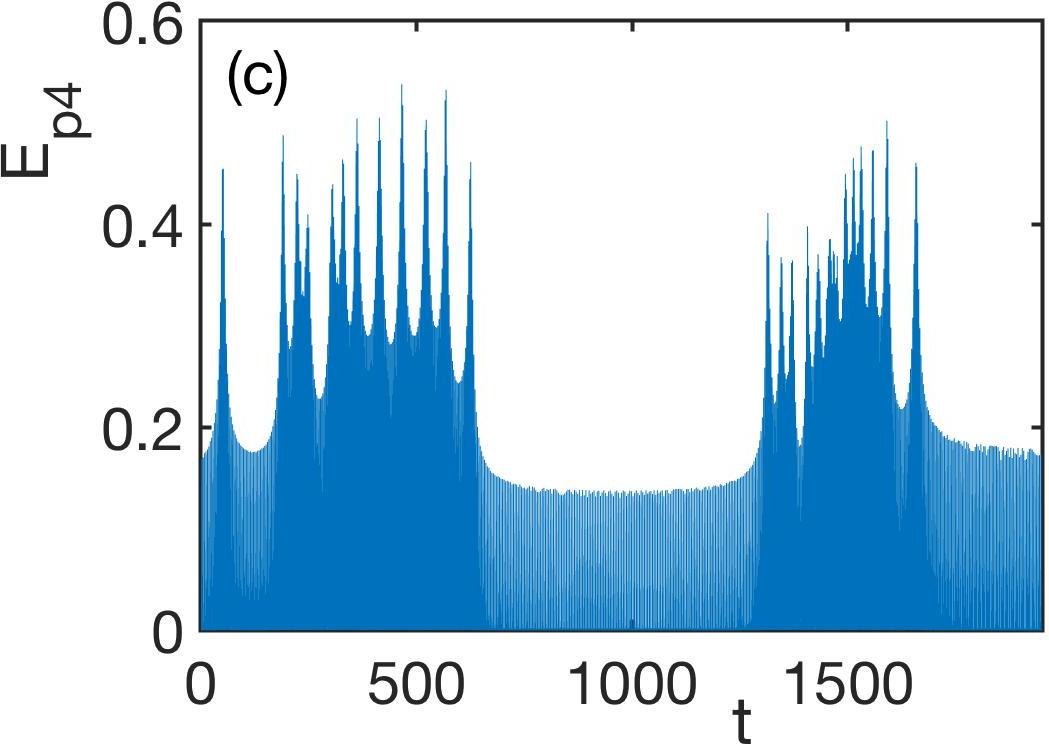} \vspace*{-0.3cm}} 
\parbox{8cm}{\includegraphics[width=7.6cm,height=6.6cm]{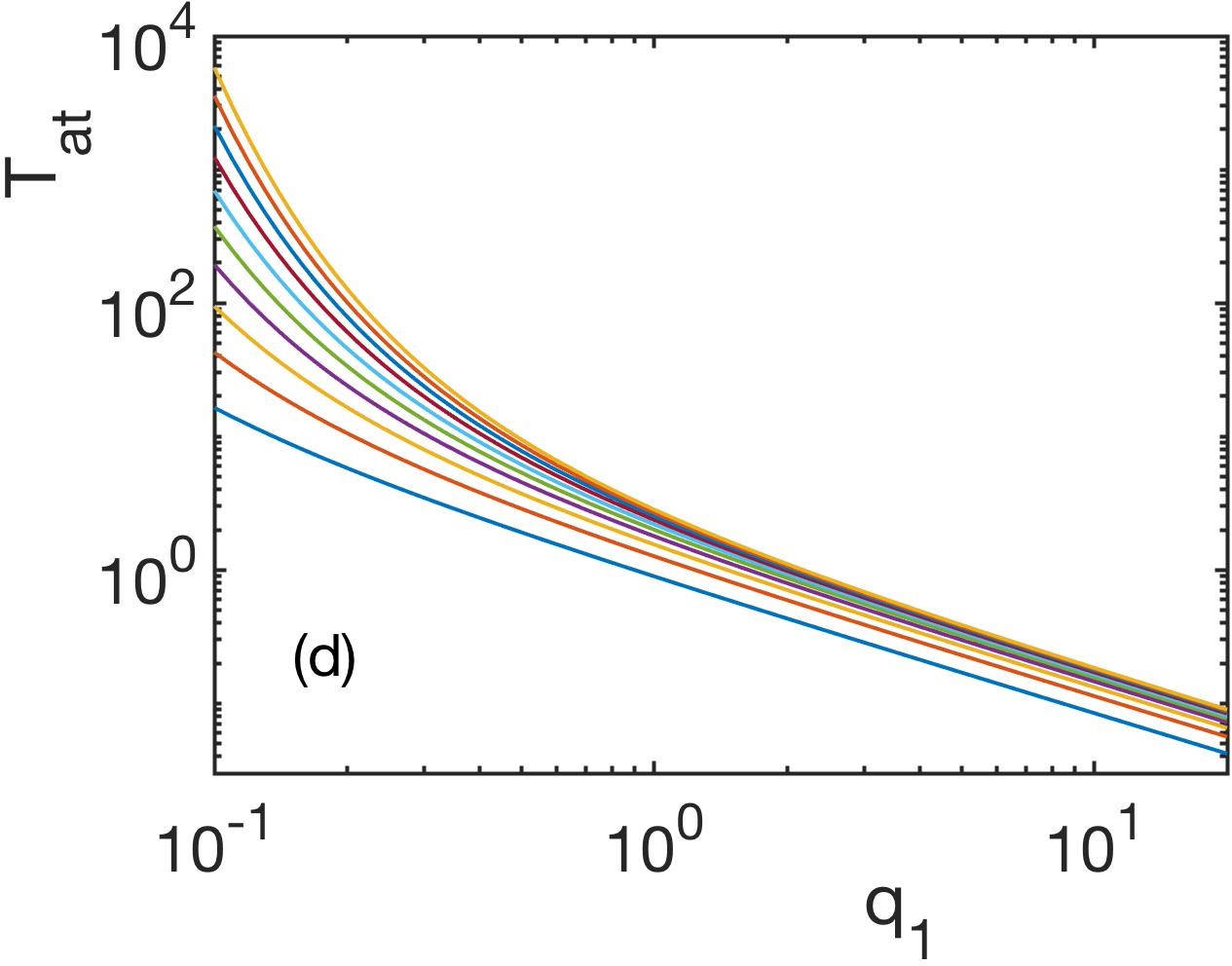}} 
\caption{(a) Time evolution $q_1(t)$ of a scattering trajectory closely approaching the saddle point
region and showing oscillations of largely different amplitudes. Initial conditions are
$q_1=30,q_i=0,i=2,...,N$ and $p_{i}=-0.53,0.65,1.08,0.58,1.92,2.15,0.35,2.57,0.53,0.06$ and the total energy is $E=8.8$. 
(b) and (c) show the specific kinetic $E_{k1}(t)$ and potential $E_{p4}(t)$ energies.
(d) The time $T_{at}$ spent above the threshold $q_2=1 \leftrightarrow E=1$ within the single
particle dynamics as a function of $q_1$. The curves from top to bottom correspond to the
energies $E=2.0,...,1.1$ in steps of $0.1$.}
\label{fig4}
\end{figure}

Let us now increase the total energy available in the scattering process for non-identical
initial conditions. A necessary condition for
further turning points to occur in the dynamics of $q_1(t)$ is (see corresponding discussion in
section \ref{dyn:lse}) the positivity of the logarithmic terms in the equation of motion (\ref{eq:heom2})
which implies that $q_i > 1$ has to occur for some particles such that the
overall sum becomes positive. Consequently certain interaction potential contributions obey $E_{pi}>1$.
Figure \ref{fig3}(b) shows for an energy $E=10.6$ the dynamics $q_1(t),(q_i(t), i=2,...,N)$
of an example trajectory with no identical ICs. Here it is clearly visible that the dof $q_1(t)$ enters in the
course of the scattering process from the CC to the saddle point region and
performs thereafter an oscillation followed by an escape back into the CC.
The dof $q_i(t),i=2,...,N$ show an increase of the amplitude of oscillations during the 
the dynamics in the saddle point region. Figure \ref{fig3}(c) shows the kinetic energy $E_{k1}(t)$
and exemplarily two of the kinetic energies $E_{ki}(t), i \ne 1$ for the same trajectory.
$E_{k1}(t)$ shows according to the oscillation of $q_1(t)$ in the saddle point region
an oscillation with three zeros. Inspecting the incoming and outgoing $E_{k1}(t),E_{ki}(t), i \ne 1$
the inelasticity of this scattering event for intermediate energies becomes visible:
$E_{k1}$ and one of the $E_{ki}$ loose energy in the course of the scattering whereas
the other $E_{ki}$ component gains energy. While this example trajectory possesses an 
energy $E>E_s$ the principal process that an oscillatory dynamics of $q_1(t)$ becomes now possible 
is by no means restricted to an energy above the saddle point energy. This is impressively
demonstrated in Figure \ref{fig4}(a,b,c).

Figure \ref{fig4}(a) shows the time evolution of $q_1(t)$ of a scattering trajectory emerging
from $q_1(t=0)=30$ and traveling towards the saddle point region. Reaching the latter we observe
a series of oscillations until, at time $t\approx 1700$, backscattering into the CC
takes place with no further turning points to occur. The many oscillations taking place possess 
very different amplitudes. Indeed, the first oscillation has its turning point at $q_1 \approx 18$
followed by a large number of oscillations with a significantly smaller amplitude. At $t \approx 600$
a huge amplitude oscillation with a turning point deep into CC is observed. Subsequently a series of
small amplitude oscillations occurs until the final escape into the CC happens. We emphasize that
such an intermittent behaviour involving backscattering and recollision events
is completely absent for the corresponding two-body system but is an inherent feature
of the many-body case. Although being a high-dimensional phase space, we could exemplarily show that
this highly oscillatory behaviour traces unstable periodic orbits which occur in the saddle point region.
This means once the trajectory gets close to one of those orbits it stays temporarily in its vicinity
i.e. it temporarily shadows the unstable periodic motion. 

A few remarks are in order. As emphasized above the corresponding two-body system shows only
simple backscattering into the CC. Scattering trajectories of the many-body system can, however, show backscattering
into the CC followed by recollision events. Once the system recollides it dwells in the saddle
point regime and finally gets backscattered into the CC. Of course, since oscillations take place
also for small amplitudes this is a crude picture of what happens indeed. According to the
analysis in section \ref{dyn:lse} a necessary condition for the occurence of a recollision event
is the surpassing of the threshold value $q_i=1$ for some dof $i$. A closer inspection reveals 
that there is generically several transversal channel dof from the set $q_i, i=2,...,N$
involved in this process: it is the sum on the r.h.s of eq.(\ref{eq:heom2}) which has to change sign in order 
to introduce the possibility of a recollision event. Indeed, the surpassing of the threshold
value leads to a deceleration
and finally a turning point in the dynamical evolution. Note that this process of repeated backscattering
and recollision does not require a fine tuning but happens generically for the regime of
intermediate energies below (and above, see next section \ref{dyn:hes}) the saddle point threshold
energy $E_s$. We remind the reader of the fact that this oscillatory behaviour is a pure dynamical
interaction effect and there is no stable equilibria of the potential landscape that would be responsible for
these processes.

To get a simple measure for the probability that our dynamical system resides in the above-threshold
regime $q_i > 1$, which enables a pronounced deceleration dynamics and finally leads to an oscillatory behaviour,
we take the following approach. We focus on the case of a single particle in the one-dimensional potential
$V(q_2;q_1)=|q_2|^{q_1}$ with a constant value for the parameter $q_1 > 0$. Assuming $q_2 > 1$ means for the
energy $E > 1$. The time which the particle spends in this regime $q_2 > 1$ in the course of a
positive half-period of its oscillation reads as follows

\begin{equation}
T_{at} = \sqrt{2} \int_{1}^{q_t} \frac{dq}{\sqrt{E-|q|^{q_1}}}
\end{equation}

for $q_t=E^{\frac{1}{q_1}}$. Figure \ref{fig4}(d) shows $T_{at}$ as a function of $q_1$ which represents the
power of the potential $V(q_2;q_1)$ for varying energy $E=1.1-2.0$ in steps of $0.1$. Obviously, $T_{at}$
is very small for large $q_1$ due to the box-like confinement and the steep walls which lead to a very
short time spent in the course of the dynamics in the region $q_2 > 1$. $T_{at}$ increases strongly
with decreasing value of $q_1$ - this increase is neither a power law nor an exponential one but of
superexponential character. It reflects the flattening of the increase of the potential $V$ for
$q_2 > 1$ in particular for $q_1 < 1$. With increasing energy the dependence of $T_{at}$ on $q_1$
becomes more pronounced. This analysis provides an intuitive explanation of the observation
that the oscillations of the trajectories of the many-body system, i.e. the backscattering
and recollision events, emanate from the saddle point region for which $q_1 < 1$ and where
the particles possess a large dwell time in the 'reactive zone' $q_i>1$.

Let us now return to our superexponential many-body system. 
Figure \ref{fig4}(b) presents the kinetic energy $E_{k1}(t)$ belonging to this heavily
oscillating scattering trajectory. We observe that small amplitude oscillations involve
high frequency energy exchange processes whereas large amplitude excursions into the CC
involve low frequency oscillations of the kinetic energy. Since small amplitude oscillations (see $q_1(t)$
in Figure \ref{fig4}(a)) are interdispersed between large amplitude oscillations we correspondingly
observe in Figure \ref{fig4}(b) bursts of high frequency kinetic energy oscillations interdispersed between
intervals of smooth variations. Correspondingly a representative of the potential energy $E_{p4}$ is
shown in Figure \ref{fig4}(c) which peaks whenever a collision with the confining walls takes place.

\subsection{High energy scattering} \label{dyn:hes}

We now turn to a discussion of the dynamics for energies above the saddle point threshold $E_s=(N-1)$.
Due to the structure of our Hamiltonian (\ref{eq:hamiltonian1}) which possesses many base dof but only
a single exponential dof $q_1$, the dynamics $q_1(t)$ determines whether backscattering into the CC
or transmission to the regions II and III of asymptotically free motion happens. Indeed, either all 
particles are backscattered or transmitted - a splitting into partial
backscattering and partial transmission is not possible.

\begin{figure}
\parbox{8cm}{\includegraphics[width=7.6cm,height=6.6cm]{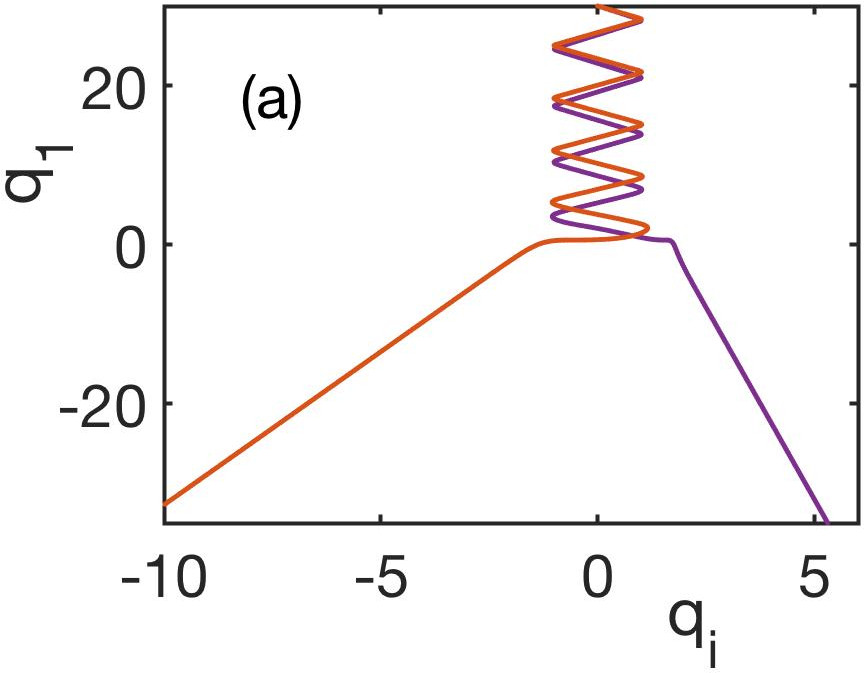}} 
\parbox{8cm}{\includegraphics[width=8.2cm,height=6.5cm]{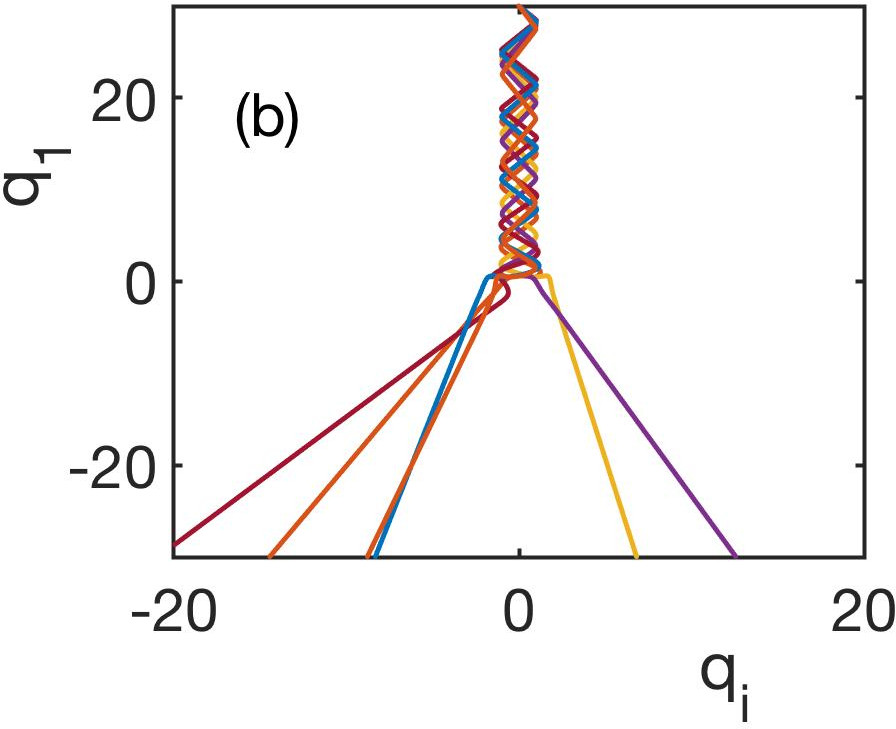}} 
\parbox{8cm}{\includegraphics[width=8.0cm,height=6.3cm]{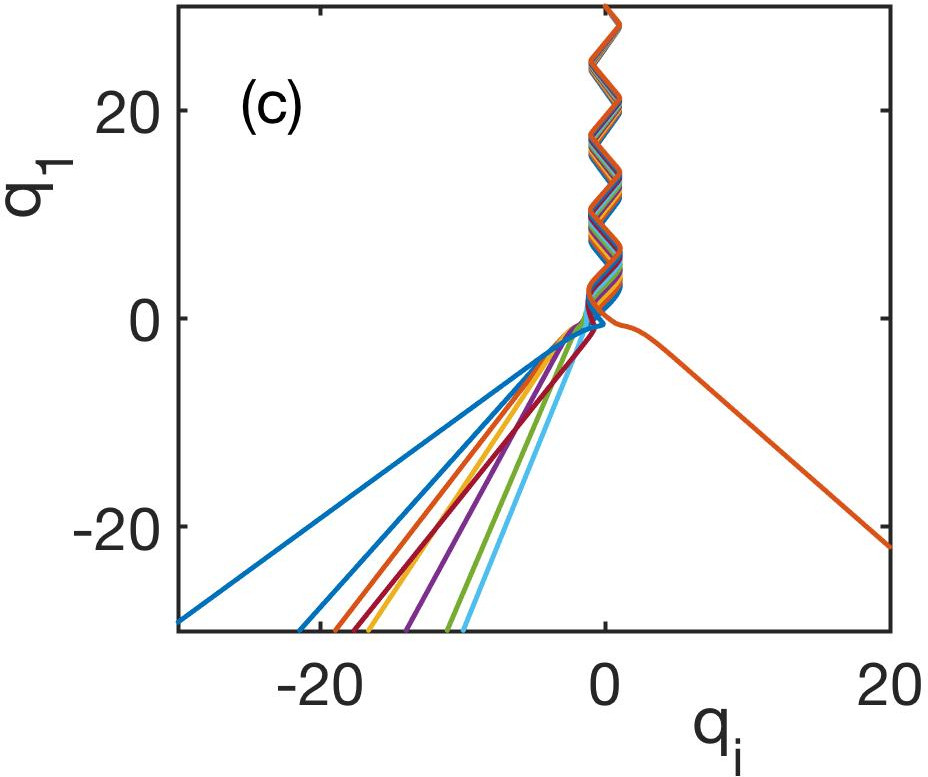}} 
\caption{Transmitting trajectories in the $(q_1,q_i)$-plane above the saddle point energy $E_s=9$ for $N=10$
particles. ICs are $q_1=30,q_i=0,i=2,...,N$ and (a) $p_1=-2.45,(p_i=1.411;i=2-5),(p_j=1.55;j=6-10)$
for a total energy $E=13$ (b) $p_i=-2.45,1.18,1.41,1.27,1.18,1.61,1.48,1.61,1.48,1.00; i=1-10$
for a total energy $E=9.1$ as well as (c) $p_i=-2.82,1.45,1.48,1.52,1.55,1.58,1.61,1.64,1.67,1.70; i=1-10$
for a total energy $E=15.2$. From (a) to (c) the distribution of the particles onto the regions
II and III of the potential landscape varies significantly.}
\label{fig5}
\end{figure}

Figure \ref{fig5}(a) shows a many-body trajectory in the $(q_1,q_i)$-planes for a total energy $E=13$, i.e.
well above the saddle point energy $E_s=9$, and for two sets of identical ICs. Consequently two
scattering processes are observed in Figure \ref{fig5}(a): the one set of identical ICs is scattered
to region II and the other set to region III (see Figure \ref{fig1}). Figure \ref{fig5}(b) shows
a trajectory in the $(q_1,q_i)$-planes for an energy $E=9.1$ slightly above the saddle point energy $E_s$ and for 
non-identical ICs except three sets of two identical ICs. In this case four scattering paths
go to the region II whereas two enter the region III while overall transmission takes place.
Finally Figure \ref{fig5}(c) shows a trajectory with energy $E=15.2$ with no identical ICs 
and as a result nine distinct paths can be observed. Eight of them go to region II and one to region III.
The above clearly demonstrates that particles can be arbitrarily distributed, after passing the saddle
point region, onto the regions II and III of asymptotic freedom. Dof with identical ICs, of course, show
identical paths.

\section{Dynamics: Statistical Properties}
\label{sec:dynamics2}

Let us now explore the statistical properties i.e. the behaviour of an ensemble of trajectories
scattering in the CC of the superexponential potential landscape. Initial conditions are $q_1=30, q_i=0, i=2,...,N$,
as in the case of the individual trajectories analyzed in the previous section, and we choose the
kinetic energies $E_{ki},i=2,...,N$ randomly from a uniform distribution with the constraint to match the energy shell. 
First we analyze the case of identical ICs for the momenta $p_{2},...,p_{N}$,
followed by the case of two sets of identical ICs and finally
the case of all ICs being different. This way the particle exchange symmetry of the Hamiltonian
is broken to an increasing extent by the chosen ICs. The main observables of our analysis are
the so-called reflection time distribution (RTD) and the momentum-time map (MTM). The reflection time
is the time interval a scattering trajectory needs to travel back to its starting-point in the CC
at $q_1=30$. The RTD represents then a histogram of the distribution of these reflection times
with varying initial conditions from the chosen ensemble. The MTM shows the intricate connection
between the initial momentum $p_1$ and the reflection time for corresponding ensembles.

\begin{figure}
\parbox{15cm}{\includegraphics[width=15cm,height=6.5cm]{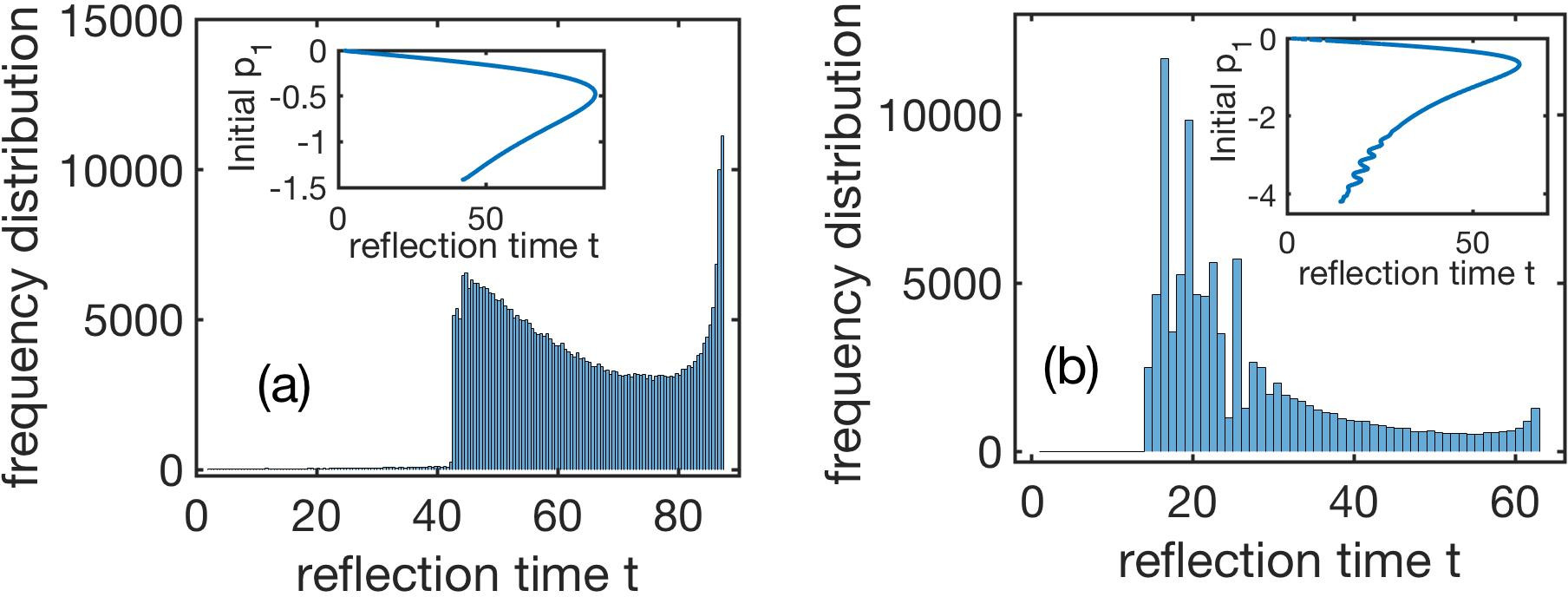}} 
\caption{Reflection time distribution for scattering in the CC (region I).
ICs are $q_1(t=0)=30,q_i(t=0)=0$. All further ICs for $p_i,i=2,...,N$ are identical. (a) Parameters
are $E=1,N=10$. The ensemble consists of $4 \cdot 10^{5}$ trajectories with randomly chosen kinetic energies.
Inset: The corresponding momentum-time map which provides the initial momentum $p_1$
versus the reflection time. (b) $E=8.8$ and the ensemble consists of $10^5$ trajectories with randomly chosen
kinetic energies. Inset: The corresponding momentum-time map.}
\label{fig6}
\end{figure}

\subsection{Ensemble properties: Identical initial conditions}
\label{sec:iic}

As discussed in section \ref{dyn:lse} the case of identical ICs w.r.t. the momenta $p_i,i=2,...,N$
for the many-body scattering dynamics
is reminescent of the corresponding behaviour of the two-body superexponential scattering dynamics 
as discussed in detail in ref.\cite{Schmelcher4}. Nevertheless, for reasons of comparison to the
generic symmetry-broken case of non-identical ICs we summarize here the main characteristics of this
case.  Figure \ref{fig6}(a,b) show the RTD and MTM for a low energy $E=1$ (a) and an energy 
$E=8.8$ (b) close to the saddle point energy $E_s=9$. The most striking observation in Figure \ref{fig6}(a)
is the appearance of two plateaus. For the first plateau given by the range $0<t\lesssim 42.5$ the typical values
of the RTD are by several orders of magnitude smaller as compared to the corresponding values in the range
$42.5 < t < 87.5$ of the second plateau. Finally a prominent peak occurs at $t \approx 87.5$.
The second plateau exhibits a broad valley towards this dominant peak. 

The origin of the above-described features of the RTD can be understood by inspecting the
corresponding MTM which is shown in the inset of Figure \ref{fig6}(a). The appearance of the
MTM, i.e. whether it is e.g. a (single-valued) curve or a spreaded point pattern, is not determined a priori.
The inset of Figure \ref{fig6}(a) shows that the MTM for the present case is a well-defined curve.
For reflection times $0 < t \lesssim 42.5$, this curve is single-valued whereas for $42.5 < t < 87.5$
it is double-valued, i.e. there appear two momentum branches of the MTM.
These two regimes correspond to the first and the second plateau of the RTD (see main figure \ref{fig6}(a)).
The time instant of the appearance of the second branch in the MTM with increasing reflection time 
is the time of the appearance of trajectories that travel to the origin of the SEP in the saddle point
region and back. The lower branch for strongly negative values of the momentum $p_2$ provides the
dominant contribution to the RTD for $t > 42.5$ providing much larger values as compared to the contribution
provided by the upper branch for $t< 42.5$. The prominent peak at $t \approx 87.5$ can be understood
by the observation that the MTM possesses at this maximal reflection time a vertical derivative: The integrated
contribution to the RTD is therefore particularly large.  This explains the overall appearance of the RTD.
For more details we refer the reader to ref.\cite{Schmelcher4}.

Figure \ref{fig6}(b) shows the RTD and MTM (see inset) for an energy $E=8.8$ close to, but still below,
the saddle point energy. For $0 < t < 13.6$ the RTD is strongly suppressed.
It shows for $t \gtrsim 13.6$ a series of peaks followed by a smooth decay up to $t \approx 63$. 
These peaks stem from the small scale oscillations present in the MTM (see inset) near the onset 
of its second branch.

\begin{figure}
\parbox{8cm}{\includegraphics[width=7.6cm,height=6.6cm]{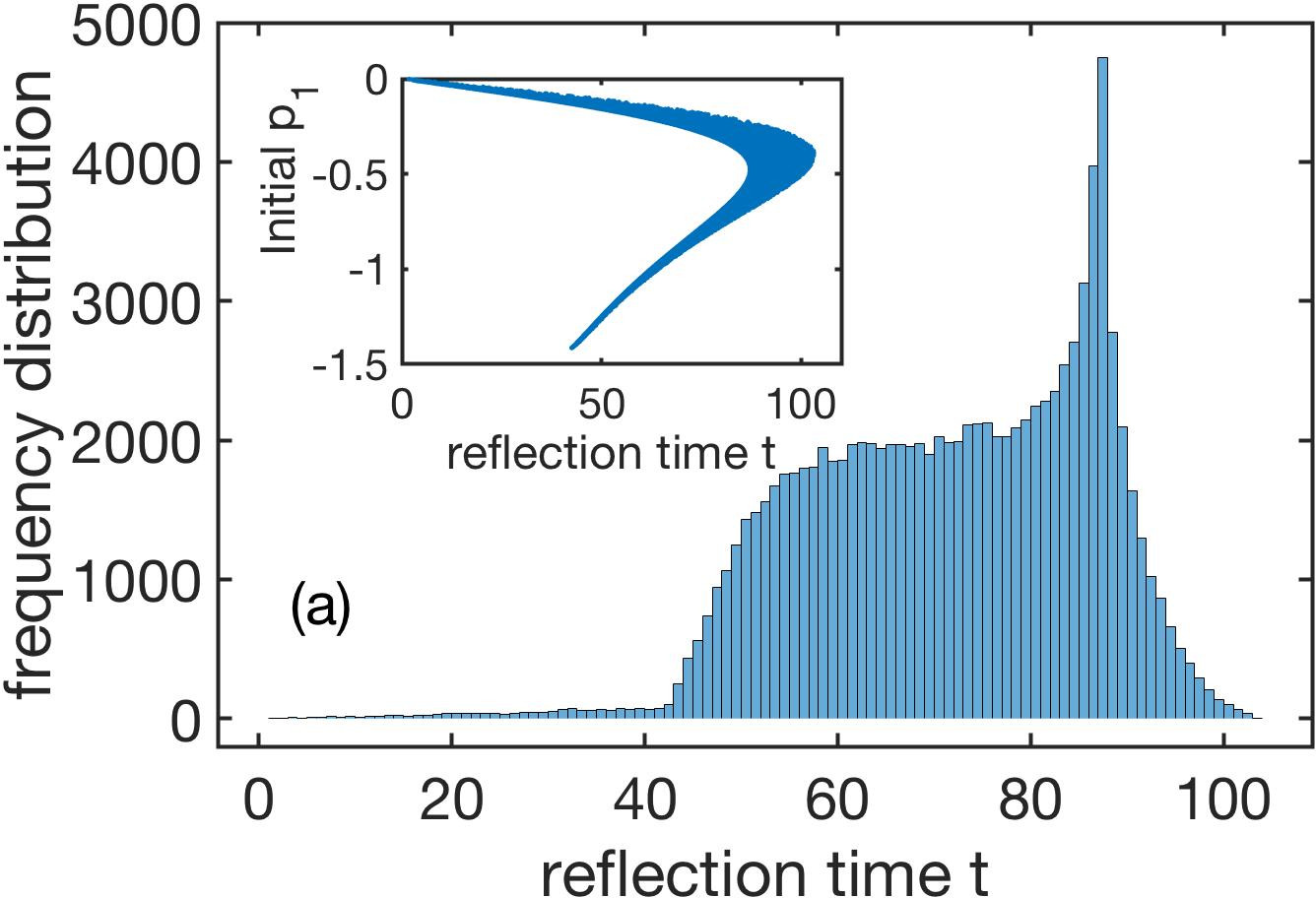}} 
\parbox{8cm}{\includegraphics[width=8.2cm,height=6.8cm]{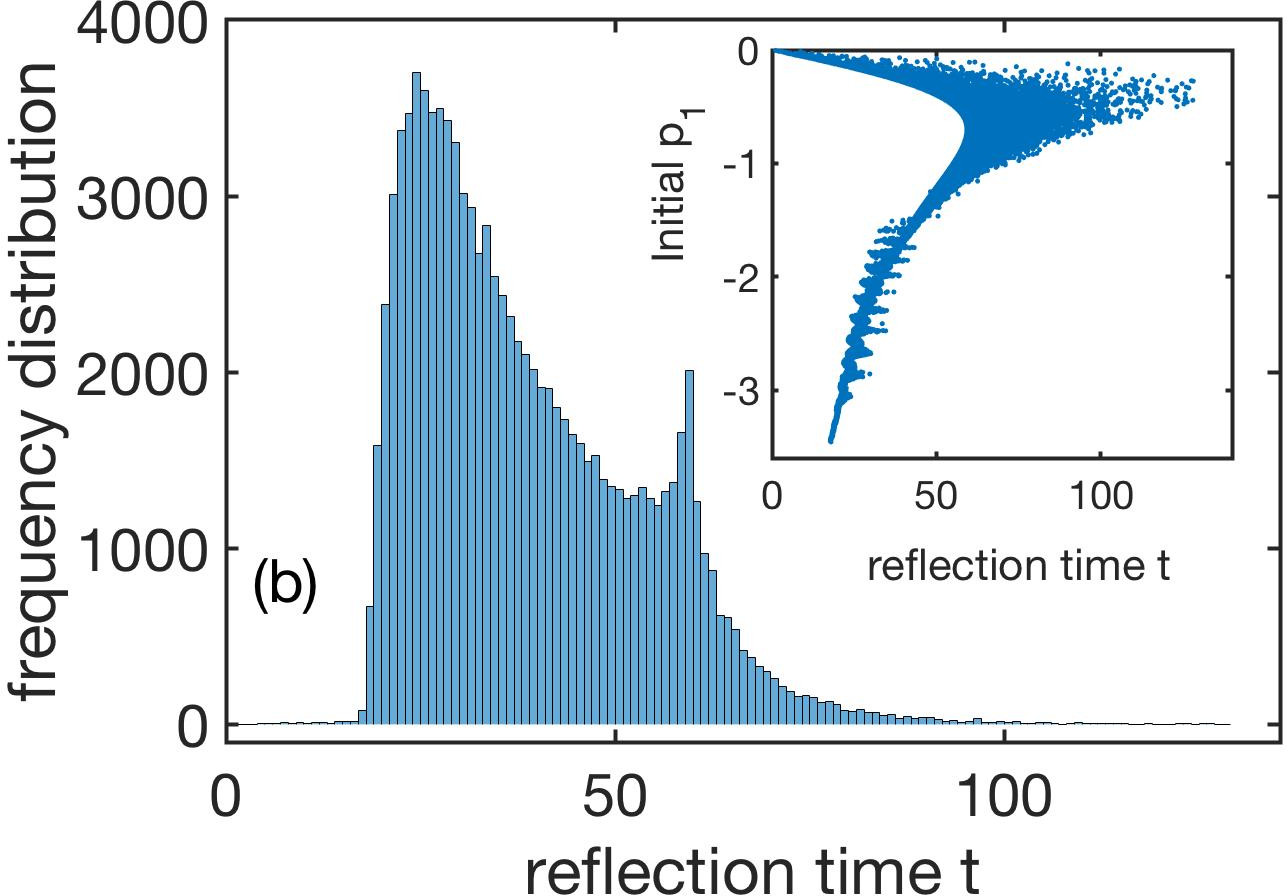}} 
\caption{Reflection time distribution for scattering in the CC (region I).
ICs are $q_1(t=0)=30,q_i(t=0)=0$. All further ICs of the dof $q_i,i=2,...,N$ belong to two classes of
identical ICs. (a) Parameters are $E=1,N=10$.
The ensemble consists of $10^{5}$ trajectories with randomly chosen kinetic energies.
Inset: The corresponding momentum-time map which provides the initial momentum $p_1$
versus the reflection time. (b) Same as (a) but for $E=6$. Inset: The corresponding momentum-time map.}
\label{fig7}
\end{figure}

\subsection{Ensemble properties: Two classes of initial conditions}
\label{sec:tsoic}

Let us now analyze the RTD for a random ensemble of trajectories that possess 
two sets of identical ICs for $p_i,i=2,...,N$, i.e. the particle exchange symmetry of the Hamiltonian
is partially broken. These ICs of the coordinates of these trajectories
obey $q_1=30,q_i=0,i=2,...,N$ as in section \ref{sec:iic}.  Figure \ref{fig7}(a) shows the RTD for an energy $E=1$.
Again two plateaus can be observed: the first one for $0<t \lesssim 43$ with a very low probability
and a second plateau for $43 \lesssim t < 104$. Opposite to the case of all identical ICs 
the increase from the first to the second plateau as well as the decrease
following the main peak is much smoother. The second plateau is essentially flat and possesses no undulation
(compare to Figure \ref{fig6}(a)). These changes can be traced back to the corresponding changes in the MTM
which is shown as an inset in Figure \ref{fig7}(a). We remind the reader that the MTM for all identical ICs concerning
$p_i,i=2,...,N$ represented a curve with two-branches (see inset of Figure \ref{fig6}(a)).
The present MTM shows a similar overall structure but the branches possess now a finite width which
increases with increasing reflection time. Again, the appearance of the second branch is responsible for
the onset of the second plateau, but now the continuous increase of the widths of the branches leads to the
observed smoothened behaviour of the RTD. Equally the smooth decay following the main peak of the RTD at $t \approx 88$
is due to the substantial extension of the MTM following the contact of the 
two distinct branches, i.e. for reflection times $t > 88$.

Figure \ref{fig7}(b) shows the RTD and MTM (see inset) for a significantly larger energy $E=6$.
Compared to the case $E=1$ (Figure \ref{fig7}(a)) a major reshaping of the RTD has taken place.
The two regions of reflection times with largely different probabilities (plateaus) are still present,
but the second plateau has become a highly asymmetric, broad and dominant peak with a maximum
at $t \approx 25$. The peak at $t \approx 60$ where the two branches of the MTM fuse (see inset of Figure \ref{fig7}(b)) 
has decreased significantly. The features of the RTD can again be interpreted in terms of the significantly
changed shape of the MTM: the onset of the second branch possesses a very steep slope and this branch exhibits
for increasing reflection times $t \gtrsim 25$ a series of 'spread transversal oscillations'. This adds up to the broad
asymmetric peak of the RTD.

\subsection{Ensemble properties: Mutually different initial conditions}
\label{sec:adic}

\begin{figure}
\parbox{16cm}{\includegraphics[width=16cm,height=7cm]{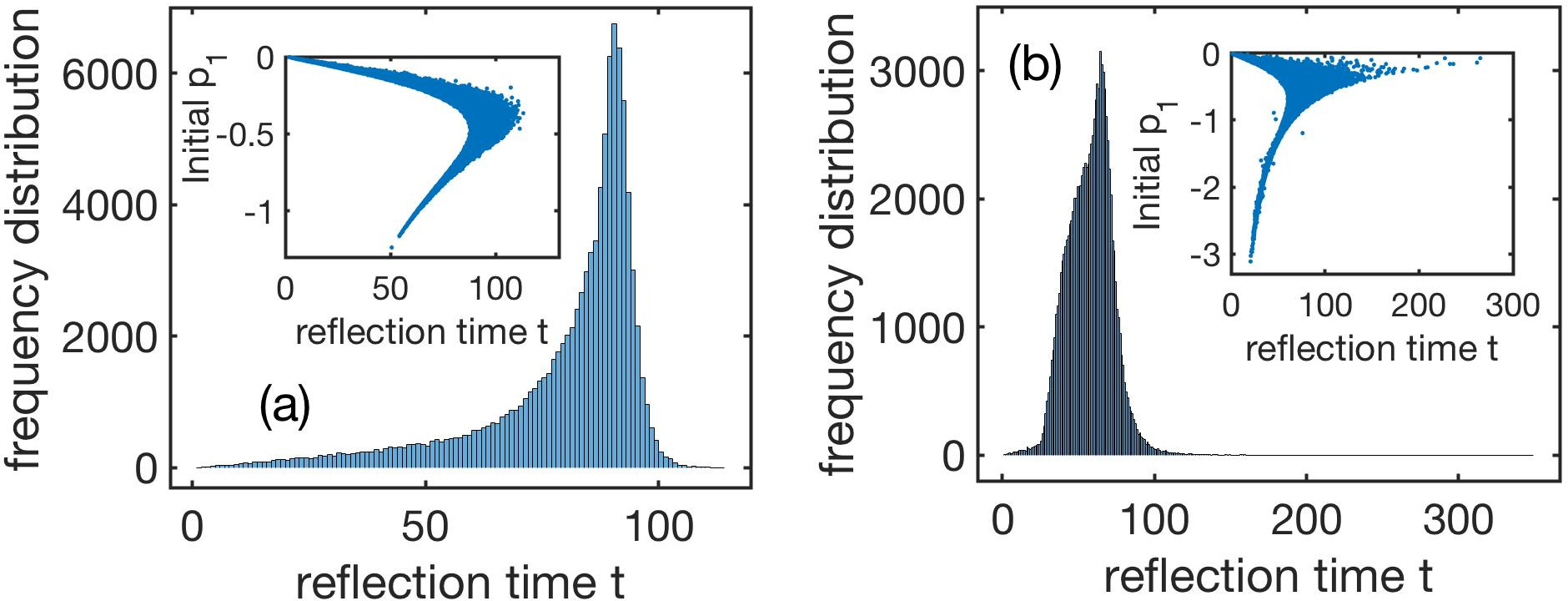}} 
\caption{Reflection time distribution for scattering in the CC (region I).
ICs are $q_1(t=0)=30,q_i(t=0)=0$. All further ICs (kinetic energies)
of the dof $q_i,i=2,...,N$ are different from each other. (a) Parameters are $E=1,N=10$.
The ensemble consists of $10^{5}$ trajectories with randomly chosen kinetic energies.
Inset: The momentum-time map. (b) Same as (a) but for $E=6$. Inset: The momentum-time map.}
\label{fig8}
\end{figure}

Let us now address the statistics of an ensemble for which all ICs of $p_i,i=2,...,N$ are different which
refers to the case of a completely broken particle exchange symmetry of the Hamiltonian.
Figure \ref{fig8}(a) shows the RTD and in the inset the corresponding MTM for an energy $E=1$.
The plateau-like structure observed in sections \ref{sec:iic} and \ref{sec:tsoic} for the unbroken
and partially exchange symmetry-broken cases, respectively, is now absent and is replaced by a single
strongly asymmetric peak centered at $t \approx 90$.
With increasing reflection times the RTD shows an accelerated increase culminating in the one central
peak while decreasing rapidly thereafter. The underlying MTM (see inset of Figure \ref{fig8}(a))
shows the typical boomerang-like structure with two broadened branches. The first branch is widening
systematically from its start at $t=0$ which is responsible for the substantial increase of the RTD for
low reflection times.

\begin{figure}
\parbox{18cm}{\includegraphics[width=17cm,height=11cm]{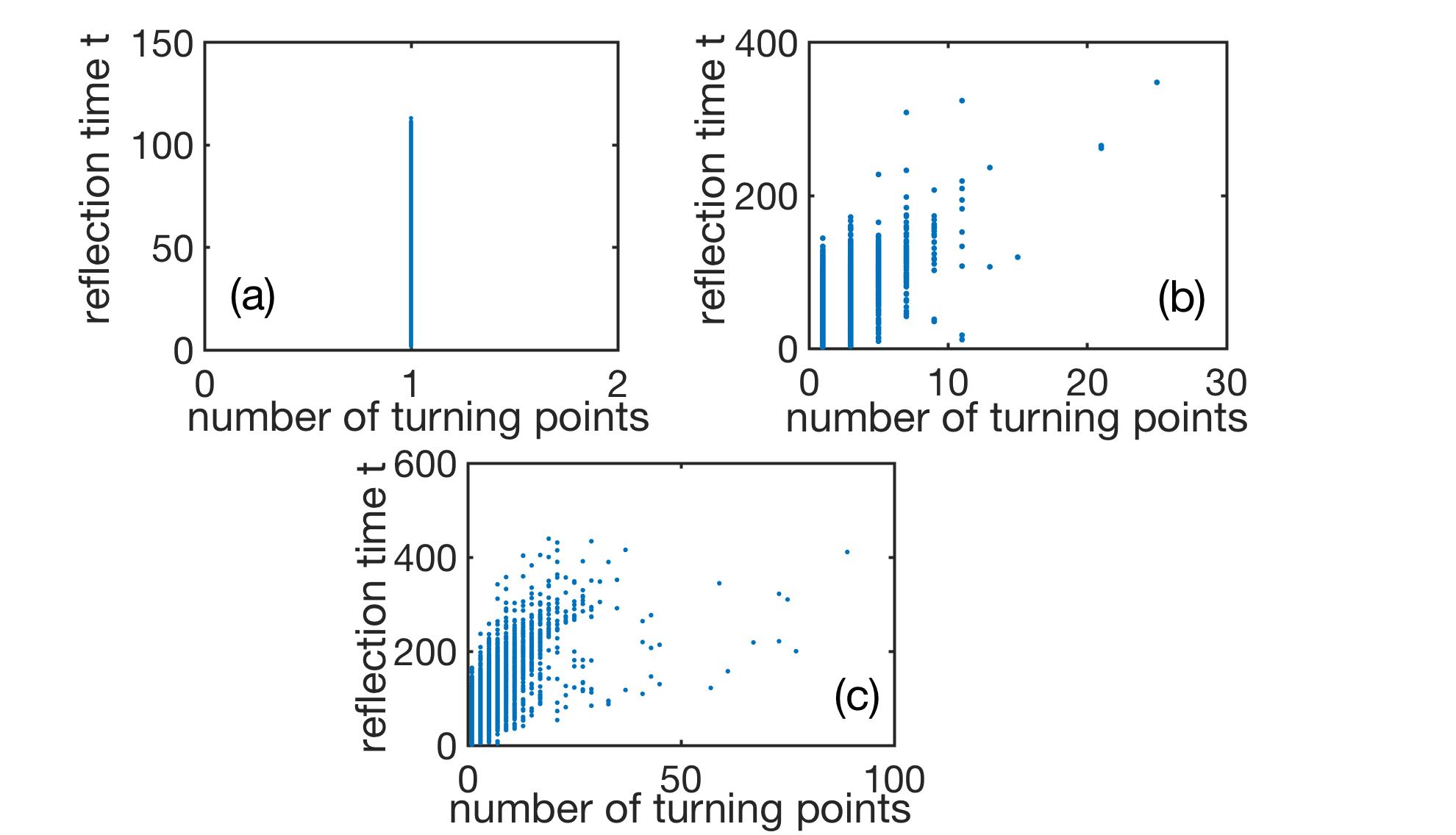}} 
\caption{Reflection time versus number of turning points of the $q_1(t)$-motion for
$E=1,6,8.8$ in (a,b,c), respectively, for an ensemble of $10^5$ trajectories. ICs
as in Figure \ref{fig8}.}
\label{fig9}
\end{figure}

Figure \ref{fig8}(b) shows the RTD and MTM for an energy $E=6$. The main differences
compared to the case $E=1$ is the reshaping of the asymmetric peak and the emergence of a very
dilute tail for large reflection times. This is reflected in the strongly distorted MTM shown
in the inset of Figure \ref{fig8}(b). The steep rise of the peak of the RTD for low reflection times
emerges again from the large slope of the second branch of the MTM. The diffuse tail of the RTD
has a corresponding counterpart in the MTM for large reflection times.

\subsection{Ensemble properties: Turning point distributions}
\label{sec:tpd}

In section \ref{sec:dynamics1} we have investigated our superexponential many-body Hamiltonian 
by analyzing the dynamics in terms of individual trajectories. The  underlying basic two-body system
\cite{Schmelcher4} shows a scattering dynamics without oscillatory behaviour w.r.t. the exponential
dof, i.e. $q_1(t)$ possesses for energies below the saddle point energy only a single turning point
which occurs at the minimal distance of the trajectories from the center of the SEP at $q_1=0$.
In section \ref{dyn:ise} we have shown that a major novelty in the many-body case is the 
oscillating structure with largely fluctuating amplitudes of trajectories experiencing the saddle
point region or physically speaking the occurrence of multiple backscattering and recollision events.
Let us now analyze the map between the reflection time of a trajectory
and its number of turning points, which we call the RTPM, for the case of mutually different IC w.r.t.
the momenta $p_i,i=2,...,N$.

Figure \ref{fig9}(a) shows the RTPM for the energy $E=1$ for the scattering dynamics in the CC.
Clearly, all trajectories and scattering events exhibit only a single turning point, and no
oscillatory dynamics is encountered. The corresponding reflection times vary continuously from zero
up to a maximal value $t \approx 114$. Increasing the energy to $E=6$ Figure \ref{fig9}(b) presents
the corresponding RTPM which shows now a large number of events up to $9$ turning points and 
a few further events up to $25$ turning points. Note, that the number of turning points is
always odd due to the fact that scattering takes place in the CC parametrized by the coordinate $q_1$.
As a general tendency one observes that the reflection time increases with the number of turning points
which is natural due to the fact that the dwell time in the saddle point region increases with increasing
number of oscillations taking place in or traversing this region. Finally Figure \ref{fig9}(c) shows
the RTPM for the energy $E=8.8$ rather close to the threshold energy $E_s=9$. As compared to the case of
$E=6$ the number of turning points possible now extends even up to approximately $90$, while the 
absolute majority of events lies below $21$ turning points. 

\section{Summary and conclusions}
\label{sec:sac}

Model systems with superexponential interaction represent a peculiar type of dynamical systems with uncommon
properties. Already for a two-body system the potential landscape shows a crossover from a confining channel (CC)
with a strongly varying transversal profile via two saddle points to a region of asymptotic freedom. The scattering
dynamics in the CC is intricate but at the same time restricted in the sense that it is a direct
in-out scattering with a single turning point of the longitudinal channel coordinate $q_1$. This situation changes
fundamentally when passing to many-body systems. In the present approach we have chosen a model system
with a single exponent degree of freedom $q_1$ for the superexponential interactions and many base degrees
of freedom $q_i,i=2,...,N$. The exponential dof $q_1$ might be considered as a 'background' or a 'guiding'
dof that determines the potential felt by the base dof. Each of the interaction terms $|q_i|^{q_{1}}$
shows the above-described geometrical crossover from channel confinement to asymptotic freedom. The many-body
Hamiltonian exhibits a particle exchange symmetry of the dof $q_i,i=2,...,N$ which can be respected, partially
broken, or completely broken by the initial conditions.

Simulating the dynamics of the many-body system we have revealed a number of important differences to the
two-body case. For low energies in the CC the $q_1(t)$ dynamics shows a transition from a step-like 
behaviour due to the spatially localized energy transfer processes to a smooth in-out scattering transition.
Increasing the energy the trajectories incoming from the CC exhibit an oscillatory behaviour emanating
from the saddle point region and possessing largely fluctuating amplitudes. This oscillatory dynamics 
comprised of backscattering and recollision events becomes increasingly more pronounced with 
increasing energy. It represent an inherent many-body effect since, generically,
all of the dof $q_i,i=2,...,N$ contribute to this process. We have analyzed this on the level of individual
trajectories but also for the case of statistical ensembles. Here the reflection time distribution shows
a characteristic transition from a two plateau structure to a single asymmetric peak behaviour. The latter
has been analyzed by inspecting the so-called momentum-time map which shows a transition from a one-dimensional
curve with two-branches to a spatially two-dimensional distribution with a characteristic shape.

There are several directions for possible future research on superexponential few- and many-body systems.
The generalization of the interaction potential to higher spatial dimensions might lead to an even
more intricate potential landscape with novel properties. The present case of a single exponent degree of
freedom and many base degrees of freedom is certainly a specific choice, and it is an intriguing perspective
to explore the case of several exponent degrees of freedom. An intriguing topic is the
statistical mechanics of our many-body system in the thermodynamical limit, where one could pose
the question whether superexponential systems relax to a stationary state of thermal equilibrium.
Finally quantum superexponentially interacting systems resulting from a canonical quantization of the many-body Hamiltonian
might show interesting scattering properties in particular due to the
squeezing channel structure and the saddle point crossover.

\section{Acknowledgments}

The author thanks F.K. Diakonos for helpful discussions and B. Liebchen for a careful reading of the
manuscript.

\end{document}